\documentclass[aps,floatfix,eqsecnum,showpacs,preprint,tightenlines]{revtex4-1}
\usepackage{epsfig}
\usepackage{bm}

\usepackage{amsmath}

\def\e{\kern+.5ex\lower.42ex\hbox{$\scriptstyle \iota$}\kern-1.10ex e}
\def\registered{{\ooalign{\hfil\raise .00ex\hbox{\scriptsize R}\hfil\crcr\mathhexbox20D}}}

\newcommand{\BA}[1]{\langle #1 \mid}

\newcommand{\KT}[1]{\mid #1 \rangle}

\begin{document}


\title{From response functions to cross sections in neutrino scattering off the deuteron and trinucleons}



\author{J. Golak}
\affiliation{M. Smoluchowski Institute of Physics, Jagiellonian University, PL-30348 Krak\'ow, Poland}
\author{R. Skibi{\'n}ski}
\affiliation{M. Smoluchowski Institute of Physics, Jagiellonian University, PL-30348 Krak\'ow, Poland}
\author{K. Topolnicki}
\affiliation{M. Smoluchowski Institute of Physics, Jagiellonian University, PL-30348 Krak\'ow, Poland}
\author{H. Wita{\l}a}
\affiliation{M. Smoluchowski Institute of Physics, Jagiellonian University, PL-30348 Krak\'ow, Poland}
\author{A. Grassi}
\affiliation{M. Smoluchowski Institute of Physics, Jagiellonian University, PL-30348 Krak\'ow, Poland}
\author{H. Kamada}
\affiliation{Department of Physics, Faculty of Engineering,
Kyushu Institute of Technology, Kitakyushu 804-8550, Japan}
\author{L.E. Marcucci}
\affiliation{Department of Physics, University of Pisa, IT-56127 Pisa, Italy
              and INFN-Pisa, IT-56127 Pisa, Italy}


\date{\today}

\begin{abstract}
Response functions, differential cross sections and total cross sections
for several (anti)neutrino induced reactions on $^2$H, $^3$He and $^3$H
are calculated in momentum space for (anti)neutrino energies up to 160 MeV,
using the AV18 nucleon-nucleon potential and a single-nucleon weak current operator.
This work is a continuation of our investigations presented in J. Golak {\it et al.}
[Phys. Rev. C{\bf 98}, 015501 (2018)].
\end{abstract}


\maketitle


\section{Introduction}
\label{section1}

Neutrinos interactions with atomic nuclei are important 
not only for nuclear physics but also for other domains like particle
physics and astrophysics. Nuclei serve as neutrino detectors in experiments 
focusing on neutrino properties, such as oscillation measurements, 
as well as in experiments where neutrinos from the interior of stars 
or from supernova explosions carry important information. 
That is why a deep understanding of neutrino induced processes 
on nuclei is necessary both for the interpretation of current 
experiments and for planning of new undertakings~\cite{Benhar2017,Alvarez-Ruso2018}.
For example, the use of the deuteron in heavy water detectors in the Sudbury 
Neutrino Observatory (SNO) for solar neutrinos 
motivated the theoretical efforts by 
Nakamura {\it et al.} \cite{PRC63.034617,NPA707.561},
Shen {\it et al.}~\cite{PRC86.035503},
Baroni and Schiavilla~\cite{Baroni17} 
to provide accurate predictions for inclusive neutrino scattering
off the deuteron. 

The results of Ref.~\cite{PRC63.034617}, a large part of the results given
in Ref.~\cite{NPA707.561} and more recent predictions by 
Shen {\it et al.}~\cite{PRC86.035503} were
obtained within the so-called "standard nuclear physics approach" 
\cite{Carlson98}, using the AV18 nucleon-nucleon (NN) force \cite{AV18} 
and augmenting the single-nucleon current
with two-nucleon (2N) contributions linked to this potential.
The latest calculations in this group, by Baroni and Schiavilla~\cite{Baroni17},
were in contrast fully based on chiral effective field theory ($\chi$EFT) 
input. The results of all these calculations performed in coordinate space
were quite similar,
which suggested that the theoretical results had a very small uncertainty
in the low-energy neutrino regime.

We could confirm these findings by performing independent calculations 
in momentum space \cite{PRC98.015501}. Namely
we investigated two- and three-nucleon
reactions with (anti)neu\-tri\-nos in the framework very 
close to the one of Ref.~\cite{PRC86.035503}
but with the single nucleon current operator.
For all the studied reactions on the deuteron 
we presented results for the total cross sections,
however, restricting ourselves to the lower (anti)neutrino energies.
We found that the few percent deviations between our strictly
nonrelativistic results 
and the predictions presented in Ref.~\cite{PRC86.035503}
originate from the relativistic kinematics, especially the phase space
factor, employed in Ref.~\cite{PRC86.035503}.
Thus our calculations for the reactions with the deuteron
passed a necessary test before embarking on three-nucleon (3N) calculations.

In Ref.~\cite{PRC98.015501} we collected important references,
which dealt with calculations for neutrino scattering on heavier 
than $A=2$ nuclei and related processes like muon capture 
or the triton beta 
decay~\cite{Gazit08,Marcucci11,Marcucci12,Marcucci13,Marcucci14,PRC90.024001,PRC94.034002,Baroni16a}. 
Here we mention only 
early calculations of the 
$\bar{\nu}_e + {^3{\rm He}} \rightarrow e^+ + {^3{\rm H}}$
and
$\bar{\nu}_\mu + {^3{\rm He}} \rightarrow \mu^+ + {^3{\rm H}}$
processes by Mintz {\it et al.}, who used an elementary particle 
model~\cite{Mintz99} dealing with non-breakup reactions, and especially work by 
Gazit {\it et al.}, who performed a number of calculations 
for neutrino induced break-up reactions with the $^3$H, $^3$He 
and $^4$He nuclei~\cite{Gazit04,Gazit07,Gazit07a},
in which final state interactions were included via the Lorentz 
integral transform method ~\cite{LITM}.
Heavier light nuclei, including $^{12}$C, were investigated 
with the Green's function Monte Carlo method~\cite{Lovato2014,Lovato2015,Lovato2018}
and using an extended factorization scheme in the spectral function formalism~\cite{Rocco2019}. 

Our calculations in Ref.~\cite{PRC98.015501} 
for the (anti)neutrino-$^3$He 
and (anti)neutrino-$^3$H inelastic scattering were limited only 
to examples of the essential nuclear response functions - we did not 
calculate any total cross sections.
In the present paper we continue our work for the (anti)neutrino 
reactions with the trinucleons. Within the same framework 
as described in Ref.~\cite{PRC98.015501} we performed several thousands 
of Faddeev calculations to gather information necessary to compute 
the differential (with respect to the lepton arm) and total 
cross sections. This information was stored in the form of the response 
functions calculated on a sufficiently dense two dimensional grid defined
by the internal nuclear energy and the magnitude of the three-momentum 
transfer. The very time consuming calculations 
of the response functions allowed us later to calculate, 
essentially in no time at all, other observables of interest. 
To this end simple two dimensional interpolations were used.
This method was of course first carefully tested for the reactions
on the deuteron, where the results of the direct calculations exist
and then applied to the trinucleons.
The results of our calculations are available upon request.

The paper is organized in the following way.
We defined all the elements of our formalism in Ref.~\cite{PRC98.015501}
so in Sec.~\ref{section2} we remind only necessary kinematical 
definitions and in Sec.~\ref{section3} we provide the connection
between the response functions and the cross sections.
In the following two sections we show selected results for
neutrino reactions on the deuteron and the trinucleons.
In particular we discuss the properties of the 2N and 3N weak 
response functions and the resulting differential and total cross sections.
Finally, Sec.~\ref{section6} contains concluding remarks and outlook.

\section{Kinematics}
\label{section2}

In Ref.~\cite{PRC98.015501} we presented our treatment of kinematics
for the (anti)neutrino induced processes with the deuteron:
$\bar{\nu}_e + {^2{\rm H}} \rightarrow e^+ + n + n$,
$\nu_e + {^2{\rm H}} \rightarrow e^- + p + p$,
$\bar{\nu}_e + {^2{\rm H}} \rightarrow \bar{\nu}_e + {^2{\rm H}}$,
$\nu_e + {^2{\rm H}} \rightarrow \nu_e + {^2{\rm H}}$,
$\bar{\nu}_e + {^2{\rm H}} \rightarrow \bar{\nu}_e + p + n$,
and
$\nu_e + {^2{\rm H}} \rightarrow \nu_e + p + n$.
Specifically in Sec.~II.A of Ref.~\cite{PRC98.015501} we discussed the differences stemming from the relativistic
and nonrelativistic formulas and introduced the internal kinetic energy
of the two-nucleon system $E_{2N}$, which entered the corresponding 
two-nucleon dynamical equations:
\begin{eqnarray} 
E_{2N} = {E} + M_d - 2 M - \sqrt{ M_e^2 + {{\bf k}^{\prime}}^{\, 2} \, }  
- \frac{ \left( {\bf k} - {\bf k}^{\prime} \, \right)^2 }{ 4 M } \, .
\label{E2N}
\end{eqnarray}
In Eq.~(\ref{E2N}) $M_d$ is the deuteron mass, the initial electron (anti)neutrino three-momentum 
is denoted by ${\bf k}$,
the final lepton three-momentum by ${\bf k}^\prime$ and its mass by $M_e$.
The energy of the initial (anti)neutrino is represented by ${E}$ and for the massless antineutrino 
$E = \mid {\bf k} \mid $. While for the charge-current (CC) 
driven reactions $M$ is equal either the proton mass $M_p$
or the neutron mass $M_n$, in the case of the neutral-current (NC) 
induced processes the ``nucleon'' mass $M=\frac12 ( M_p + M_n \, )$ and $M_e= 0$.

Our approach to the corresponding processes of $^3$He and $^3$H disintegration
is quite analogous. Since we can generally deal with a two-body and three-body 
breakup process, we outline our kinematical formulas for the case
of the CC induced disintegration of $^3$He, where a massive lepton appears 
in the final state.
Our aim is to 
calculate the maximal final lepton energy for a given lepton scattering angle $\theta$
and express the internal energy of the 3N system $E_{3N}$ in terms of $E$, $\theta$ 
and the final positron energy $E^\prime \equiv \sqrt{ M_e^2 + {{\bf k}^{\prime}}^{\, 2} \, }$.
This information is necessary to evaluate the differential and total cross sections 
for the (anti)neutrino induced $^3$He and $^3$H disintegration processes.

We start with the two-body breakup of $^3$He,
$\bar{\nu}_e + {^3{\rm He}} \rightarrow e^+ + n + d$, and
the exact relativistic form of the energy and momentum conservation in the laboratory frame:
\begin{eqnarray} 
{E} + {M_{^3{\rm He}}} & = &
\sqrt{ M_e^2 + {{\bf k}^{\prime}}^{\, 2} \, }  
+
\sqrt{ M_n^2 + {\bf p}_{n}^{\, 2} \, }
+ 
\sqrt{ M_d^2 + {\bf p}_{d}^{\, 2} \, }  \, , \nonumber  \\
{\bf k} &=& {\bf k}^\prime + {\bf p}_n + {\bf p}_d \, ,
\label{relnd}
\end{eqnarray}  
where ${\bf p}_n$ and ${\bf p}_d$ stand for the final neutron and deuteron 
momenta, respectively.

The maximal energy of the emerging lepton under a given
scattering angle $\theta$, where $\cos\theta = \hat{\bf k} \cdot {\hat{\bf k}}^{\, \prime}$,
can be obtained by considering the square of the total four momentum of the nuclear system $s_{nuc}$,
\begin{eqnarray}
s_{nuc} \equiv
\left( \, \sqrt{ M_n^2 + {{\bf p}_{n}}^{\, 2} \, }  
+ 
\sqrt{ M_d^2 + {{\bf p}_{d}}^{\, 2} \, }  \, \right)^2 - \left( {\bf p}_{n} + {\bf p}_{d} \, \right)^2 \, .
\label{snuc}
\end{eqnarray}

Note that as for the reactions with the deuteron there is no restriction on the scattering angle $\theta$.
The invariant quantity $s_{nuc} $ can be evaluated in any reference frame and through the four 
momentum conservation expressed in terms of the lepton momenta 
${\bf k} $ and ${\bf k}^\prime$.  Considering 
a system, where the total three momentum of the neutron-deuteron system is zero, we find that 
\begin{eqnarray}
s_{nuc} = \left( {E} + {M_{^3{\rm He}}} - E^\prime \, \right)^2 - 
\left(  {\bf k} - {\bf k}^\prime \, \right)^2 \ge \left( M_n + M_d \, \right)^2 \, .
\label{snuc.2}
\end{eqnarray}
The condition for $(E^\prime)_{max}^{nd}$ in the two-body breakup case reads thus 
\begin{eqnarray}
s_{nuc} = \left( M_n + M_d \, \right)^2 
\label{snuc.3}
\end{eqnarray}
and can be obtained analytically, providing the reference result for the 
approximation discussed below. 
The corresponding equation for the maximal 
positron energy in the three-body breakup case, $(E^\prime)_{max}^{nnp}$, 
has a similar form, namely 
\begin{eqnarray}
s_{nuc} = 
\left( 2 M_n + M_p \, \right)^2 \, .
\label{snuc.4}
\end{eqnarray}
The relativistic expression for the two reactions can be written in the following form
\begin{eqnarray}
(E^\prime)_{max} = \frac12 \,  \bigg( -E^2 E_{ini} + E_{ini}^3 + E_{ini} M_e^2 - E_{ini} M_{tot}^2 +  \nonumber \\
E \cos\theta \Big( E^4 + E_{ini}^4 + \left( M_e^2 - M_{tot}^2 \right)^2  \nonumber \\
- 2 E^2 \left( E_{ini}^2 + M_e^2 - 2 M_e^2 \cos^2 \theta - M_{tot}^2 \right)  \nonumber \\
-2 E_{ini}^2 \left( M_e^2 + M_{tot}^2 \right) \Big)^{\frac12} \, \bigg) \, 
\frac1{ E_{ini}^2 - E^2 \cos^2 \theta \, } \, ,
\label{eprimemaxrel}
\end{eqnarray}
where $E_{ini} \equiv E + {M_{^3{\rm He}}}$ and $M_{tot} \equiv M_n + M_d$ 
($M_{tot} \equiv 2 M_n + M_p$) for the two-body (three-body) breakup reaction.

Since our dynamical equations are solved within a nonrelativistic framework,
employing some further kinematical simplifications,
we prepare a consistent set of kinematical conditions. We start again with 
the energy conservation, using nonrelativistic formulas in the nuclear sector
\begin{eqnarray} 
{E} + {M_{^3{\rm He}}} & = &
E^\prime
+ 
M_n + M_d +
\frac { {\bf p}_n^{\, 2} \, }  { 2 M_n}  
+
\frac { {\bf p}_d^{\, 2} \, }  { 2 M_d} \, \nonumber \\
& \approx &
E^\prime 
+ 
3 M - \mid {B_{^2{\rm H}}} \mid +
\frac { {\bf p}_n^{\, 2} \, }  { 2 M}  
+
\frac { {\bf p}_d^{\, 2} \, }  { 4 M} \, ,
\label{nrlnd}
\end{eqnarray}
where $M$ is the ``nucleon'' mass and ${B_{^2{\rm H}}}$ is the deuteron binding energy.
We require that the kinetic energy of the 
nuclear system in its total momentum zero frame $E_{nd}$ must be  
non-negative. With 
${M_{^3{\rm He}}} = 3 M - \mid {B_{^3{\rm He}}} \mid $, where ${B_{^3{\rm He}}}$ is the $^3$He binding energy,
one obtains:
\begin{eqnarray} 
E_{nd} = E - E^\prime - \mid {B_{^3{\rm He}}} \mid + \mid {B_{^2{\rm H}}} \mid 
- \frac{ \left( {\bf k} - {\bf k}^{\prime} \, \right)^2 }{ 6 M } \ge 0  \, .
\label{nrlnn71}
\end{eqnarray}
Similarly, for the full breakup we find
\begin{eqnarray} 
E_{3N} = E - E^\prime - \mid {B_{^3{\rm He}}} \mid 
- \frac{ \left( {\bf k} - {\bf k}^{\prime} \, \right)^2 }{ 6 M } \ge 0  \, .
\label{nrlnn72}
\end{eqnarray} 
Here we have introduced also the three-nucleon internal energy $E_{3N}$, which
will be used 
together with the magnitude of the three momentum transfer $ Q \equiv \mid  {\bf k} - {\bf k}^{\prime} \mid$
to label the nuclear response functions.
The nonrelativistic values of the maximal positron energies in the final state  
$(E^\prime)_{max}^{nd}$,
and
$(E^\prime)_{max}^{nnp}$,
can be obtained from the conditions 
$ E_{nd} = 0 $ and $ E_{3N} = 0$, respectively.
Each of these two conditions can be cast in the form of
a forth degree equation, which we chose to solve numerically.
As a starting value in the numerical search one can equally well take
two analytically known results: the relativistic expressions from Eq.~(\ref{eprimemaxrel}) or the 
nonrelativistic formulas stemming from Eqs.~(\ref{nrlnn71})--(\ref{nrlnn72}) with $M_e = 0$:
\begin{eqnarray} 
(E^\prime)_{max}^{nd} & \approx &
\sqrt{9 M^2 + 6 M (\mid {B_{^2{\rm H}}} \mid  - \mid {B_{^3{\rm He}}} \mid + E - E \cos\theta ) - E^2 \sin^2 \theta } + E \cos\theta -3 M \, , \nonumber \\
(E^\prime)_{max}^{nnp} & \approx &
\sqrt{9 M^2 + 6 M ( - \mid {B_{^3{\rm He}}} \mid + E - E \cos\theta ) - E^2 \sin^2 \theta } + E \cos\theta -3 M \, .
\label{nrlnn8}
\end{eqnarray}
We compared the results for the exact relativistic formulas with
their approximate nonrelativistic analogues and found 
that for up to $E \le$ 160 MeV the maximal relative difference 
between these values did not reach 1~\% for the whole allowed 
range of $\theta$ angles.

It is clear that the kinematics of the other processes of (anti)neutrino induced
breakup of the trinucleons can be analyzed in the same way. In particular, for
the NC driven reactions, where massless neutrinos appear in the final state,
the analytical formulas from Eq.~(\ref{nrlnn8}), 
derived for the nonrelativistic approximation we employ in the paper, become exact.

\section{The cross sections and response functions}
\label{section3}

The formalism of neutrino scattering off nuclei is well established,
see for example \cite{Walecka}. 
For the CC induced processes it stems directly from the Fermi theory
but it has to be modified to include additionally the NC based processes.
For the lowest order processes 
the transition matrix element can be written as a contraction of the 
nuclear part $N^\lambda$ and 
the leptonic part $L_\lambda$, where the latter is expressed 
in terms of the Dirac spinors and gamma matrices so we can focus on matrix elements
\begin{eqnarray}
N^\lambda = 
\BA{\Psi_f \, {\bf P}_f \, m_{f} \, } \, 
j_{W}^\lambda
\, \KT{\Psi_i \, {\bf P}_i \, m_{i} \, } 
\label{nlambda}
\end{eqnarray}
of the nuclear weak charged or neutral current
$j_{W}^\lambda$ between the initial 
$ \KT{\Psi_i \, }$ 
and final 
$ \KT{\Psi_f \, }$ 
nuclear states, where
the total initial (final) nuclear three-momentum is denoted by ${\bf P}_i$ (${\bf P}_f$),
$m_i$ is the initial nucleus spin projection and $m_f$ is the set of spin projections
in the final state. As explained in Ref.~\cite{PRC98.015501} it is advantageous 
to assume a system of coordinates, where $ {\bf Q} \equiv {\bf k}  -  {\bf k}^{\, \prime} \parallel {\hat z}$
and ${\hat y} = \frac{ {\bf k} \times  {\bf k}^{\, \prime} } {\mid  {\bf k} \times  {\bf k}^{\, \prime} \mid }$, so
\begin{eqnarray}
k^\prime_x & = & k_x  =  \mid  {\bf k} \mid \mid  {\bf k}^{\, \prime} \mid \sin \theta / \mid  {\bf Q} \mid \, , \nonumber \\
k^\prime_y & = & k_y  =  0 \, , \nonumber \\
k_z & = & \mid  {\bf k} \mid \left( \mid  {\bf k} \mid  - \mid  {\bf k}^{\, \prime} \mid \cos \theta \, \right)  / \mid  {\bf Q} \mid \, , \nonumber \\
k^\prime_z & = & \mid  {\bf k}^{\, \prime}  \mid \left( -\mid  {\bf k}^{\, \prime}  \mid  + \mid  {\bf k} \mid \cos \theta \, \right)  / \mid  {\bf Q} \mid \, , \nonumber \\
\mid  {\bf Q} \mid & = & \sqrt{ {\bf k}^{\, 2} +  {{\bf k}^{\, \prime \, 2 }} - 2 \mid  {\bf k} \mid  \mid  {\bf k}^{\, \prime} \mid \cos \theta \, } \, .
\label{kinemtheta}
\end{eqnarray}

Further, the essential part of the square of the transition matrix element 
$ \mid L_\lambda N^\lambda \mid ^2 $ can be written as 
\begin{eqnarray}
\mid L_\lambda N^\lambda \mid ^2 & = &
V_{00} \mid N^0 \mid^2 \, + \, 
V_{MM} \mid N_{-1} \mid^2 \, + \, 
V_{PP} \mid N_{+1} \mid^2 \, \nonumber \\
& + &  V_{ZZ} \mid N_{z} \mid^2 \, + \, 
V_{Z0} \, {\rm Re} \big( 2 \, N_{z} \, \left( N^0 \right)^* \, \big)  \, ,
\label{Tsquared}
\end{eqnarray}
where the $V_{ij}$ functions arise from the leptonic arm
and the spherical components are used for $N^\lambda$~\cite{PRC98.015501}.
By ``essential part'' we mean the part, which contributes 
to the cross sections 
in the case, where information about the nuclear sector is integrated over 
and only the final lepton momentum is known. We restrict ourselves only to such cases
in the present work.
For the neutrino induced reactions we get
\begin{eqnarray}
V_{00} & = & 8 \, \left(  {\bf k}^{\, \prime} \cdot {\bf k} + E \, E^\prime \, \right) \, , \nonumber \\
V_{MM} & = & 8 \, \left( E + k_z \, \right) \, \left( E^\prime - k^\prime_z \, \right) \, , \nonumber \\
V_{PP} & = & 8 \, \left( E - k_z \, \right) \, \left( E^\prime + k^\prime_z \, \right) \, , \nonumber \\
V_{ZZ} & = & 8 \, \left(  -{\bf k}^{\, \prime} \cdot {\bf k} + E \, E^\prime + 2 k_z k^\prime_z  \, \right) \, , \nonumber \\
V_{Z0} & = & -8 \, \left( E \, k^\prime_z  + E^\prime \, k_z \, \right) 
\label{V-fun}
\end{eqnarray}
and for reactions with antineutrinos the
corresponding $\bar{V}_{ij}$ functions are easily obtained from the previous set:
\begin{eqnarray}
\bar{V}_{00} & = & V_{00}  \, , \nonumber \\
\bar{V}_{MM} & = & V_{PP}  \, , \nonumber \\
\bar{V}_{PP} & = & V_{MM}  \, , \nonumber \\
\bar{V}_{ZZ} & = & V_{ZZ}  \, , \nonumber \\
\bar{V}_{Z0} & = & V_{Z0}  \, .
\label{BARV-fun}
\end{eqnarray}

The standard steps, which take also into account the normalization 
of the Dirac spinors and nuclear states,
lead to the final form of the cross section.
For the CC neutrino induced reactions we obtain
\begin{eqnarray}
\frac{d^{\, 3} \sigma }{ dE^\prime \, d{\Omega}^\prime } \, = \,
\frac{ G_F^2 \cos^2\theta_C} { \left( 2 \pi \right)^2 } \, 
F(Z, E^\prime) \, 
\frac{\mid {\bf k}^{\, \prime} \mid}{ 8 E } \, 
\Big(
V_{00} R_{00,CC} + 
V_{MM} R_{MM,CC} + 
\nonumber \\
V_{PP} R_{PP,CC} + 
V_{ZZ} R_{ZZ,CC} + 
V_{Z0} R_{Z0,CC}
\, \Big) \, ,
\label{dsigma3}
\end{eqnarray}
where the value of the Fermi constant, $G_F = 1.1803 \times 10^{-5} \, {\rm GeV}^{-2} $,
and $\cos\theta_C = 0.97425$ are taken from Ref.~\cite{PRC86.035503}. 
The Fermi function $F(Z, E^\prime)$ \cite{Fermi34} is introduced to account for the
distortion of the final lepton wave function by its Coulomb interaction 
with more than one proton in the final state and is not needed otherwise.
For the NC driven processes the Fermi function $F(Z, E^\prime) $ and
$ \cos^2\theta_C$ are dropped in Eq.~(\ref{dsigma3}).
Our predictions for all the CC induced reactions are valid only for the 
electron flavor but for the NC reactions results are the same for all the three flavors.   

The essential dynamical ingredients in the inclusive cross section are the nuclear response functions
originating from the integration of
various products of the nuclear matrix elements
over the whole nuclear phase space available for the fixed final 
lepton momentum:
\begin{eqnarray}
R_{AB,CC}  = \sum\limits_{m_i,m_f} \int df \,
\delta \left( E_{3N} - E_f  \right) \,
  \langle \Psi_{f} \mid j^{A}_{3N} \mid \Psi_i \rangle \,
\left( \langle \Psi_{f} \mid j^{B}_{3N} \mid \Psi_i \rangle \right)^* \, ,
\label{RAB}
\end{eqnarray}
where $AB$= $00$, $MM$, $PP$, $ZZ$ and $Z0$,
$m_i$ and $m_f$ represent the whole sets
of the initial and final spin magnetic quantum numbers, respectively, while
the $df$ integral denotes the sum and the integration over all final
3N states with the fixed energy $E_{3N}$. 
The 3N bound state wave functions are calculated using the method
described in Ref.~\cite{Nogga.1997}. 
The direct integration would allow one to evaluate 
contributions from any part of the phase space. In particular, for the reactions 
on $^3$He and $^3$H it would be possible to obtain contributions from the two- 
and three-body breakup channels. However, the numerical cost of such calculations 
needed for the total cross section, which is the main objective of the present paper
is very high. Thus we decided to compute the 3N response functions in a much more economical way,
using closure and employing the special Faddeev
scheme \cite{incRAB,raport2005}. In Ref.~\cite{PRC98.015501} we compared results based 
on these two quite different approaches and obtained a very good agreement.
The results for all the 3N response functions presented in this paper 
are obtained with the second, cheaper method. This closure-based scheme 
could be formulated also for 
the 2N system but in that case the integration over the phase space 
is well under control and perfectly practical as will be explained in the following.
Thus all the 2N response functions are obtained as in Ref.~\cite{PRC98.015501}, 
by direct integrations.

It is very important to realize that while the differential cross section
$\frac{d^{\, 3} \sigma }{ dE^\prime \, d{\Omega}^\prime }$ depends on three 
kinematical variables, $E$, $\theta$ and $E^\prime$, the response functions 
are defined in terms of the internal nuclear energy ($E_{2N}$ or $E_{3N}$) 
and the magnitude of the three momentum transfer ($Q$). 
We will use this feature to facilitate the calculations.

Before we discuss our results, we remind the reader of the most important features 
of our momentum space framework.
We follow the path paved by Refs.~\cite{PRC63.034617,NPA707.561,PRC86.035503,Baroni17}
whose authors investigated inclusive neutrino scattering on the
deuteron with configuration space methods. Those very advanced investigations were
performed with traditional and chiral NN potentials and included weak nuclear current 
operators with a one-body part and two-body contributions, adjusted to the NN force.
Here we continue our work from Ref.~\cite{PRC98.015501} 
with the standard AV18 NN potential \cite{AV18} and the single nucleon current 
operator defined in Ref.~\cite{PRC90.024001}.
This form and parametrization of the single nucleon current was previously 
employed for example in Refs.~\cite{Marcucci11,PRC90.024001}.
Since we restrict ourselves to the low (anti)neutrino energies, $E \le 160$~MeV, 
where our nonrelativistic approach is fully justified, we expect, based on the 
results of Ref.~\cite{PRC86.035503}, that 2N contributions in the current operators 
would lead to effects smaller than $2-4 \%$.

\section{Results for (anti)neutrino scattering on $^2$H}
\label{section4}

\begin{figure}
\includegraphics[width=0.31\textwidth]{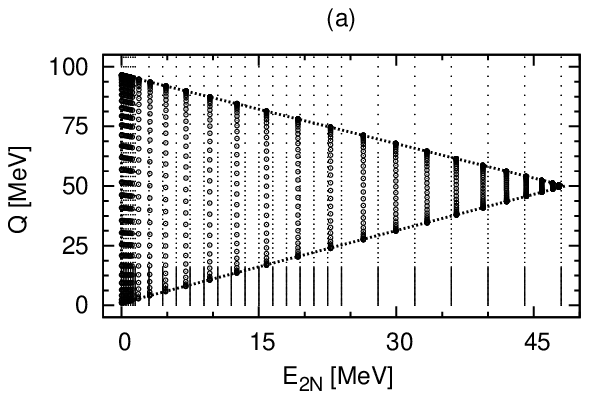}
\includegraphics[width=0.31\textwidth]{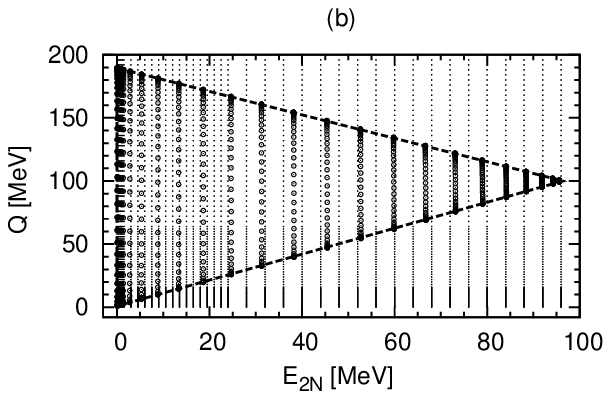}
\includegraphics[width=0.31\textwidth]{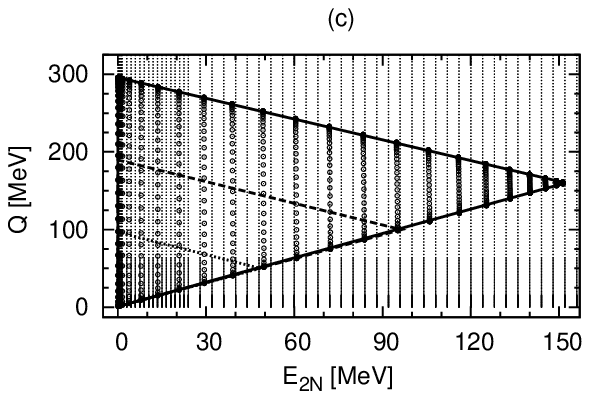}
\caption{The rectilinear grid of $(E_{2N},Q)$ points used to store the response
functions for the ${\nu}_e + {^2{\rm H}} \rightarrow e^- + p + p$ reaction (tiny dots) 
and the actual $(E_{2N},Q)$ points used to evaluate the total cross section (circles)
in the triangle-like area 
for the initial neutrino energy $E$= 50 MeV (a), 100 MeV (b) and 160 MeV (c).
Lines, which separate the physical region for a given $E$ are obtained 
from approximate Eqs.~(\ref{Qmin})-(\ref{E2NMAX}). For $E$= 160 MeV the 
border lines for the two smaller energies are also shown.
}
\label{FIG.CC-NUE-2H.RANGES}
\end{figure}

Calculating the total cross section for (anti)neutrino induced breakup reactions 
for many initial (anti)neutrino energies, starting for each energy anew,
could lead in fact to a waste of computer resources. As we show in Fig.~\ref{FIG.CC-NUE-2H.RANGES}
for the ${\nu}_e + {^2{\rm H}} \rightarrow e^- + p + p$ reaction,
while calculating the total cross section for increasing initial antineutrino energies, 
the dynamical information in the form of response functions is taken from
a part of the $(E_{2N},Q)$ domain, which necessarily overlaps with the region
corresponding to lower energies. 
This is also true for the reactions with the trinucleons.
That is why calculating response functions
on a sufficiently dense grid and using stored values for interpolations
to integral $(E_{2N},Q)$-points appears to be advantageous. 
The same stored response functions can be used not only to generate 
the total cross sections but also to calculate the intermediate differential cross sections
$d^3{\sigma}/\left( dE^\prime \, d{\Omega}^\prime \, \right)$ 
and 
$d{\sigma}/d{\theta} $.
Yet another advantage of storing response functions becomes clear for the NC induced 
processes, where {\em the same} response functions are used for the neutrino and 
antineutrino induced reactions.

Before embarking on 3N calculations, this approach 
was tested in the 2N system, where results of the direct calculations of the total cross sections
as defined for example in Eq.~(2.22) of Ref.~\cite{PRC98.015501} and predictions based on the response 
functions' interpolations could be easily compared.  
The various sets of response functions should be prepared with great care, taking 
into account the character of their dependence on $E_{2N}$ and $Q$.
We decided to use simple rectilinear grids, which forced us to use many points on the whole grid, 
even if a sharp maximum was strongly localized, leading however to very accurate predictions.
This feature of our calculations is clearly visible in Fig.~\ref{FIG.CC-NUE-2H.RANGES}.

\begin{figure}
\includegraphics[width=0.45\textwidth]{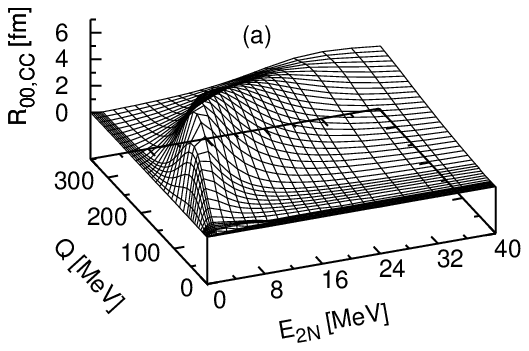}
\includegraphics[width=0.45\textwidth]{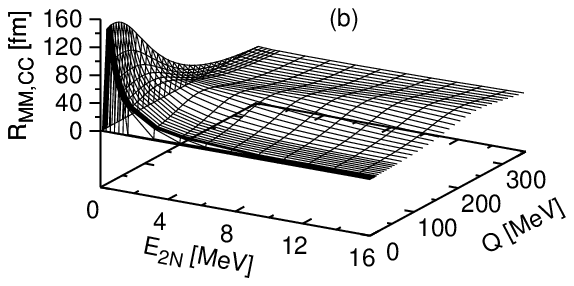}
\includegraphics[width=0.45\textwidth]{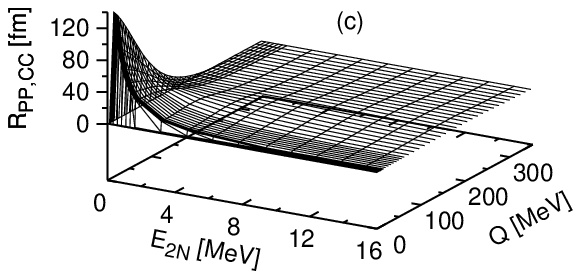}
\includegraphics[width=0.45\textwidth]{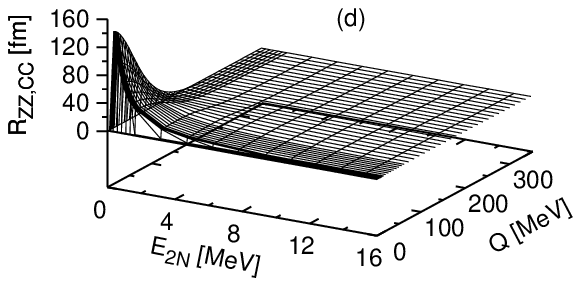}
\includegraphics[width=0.45\textwidth]{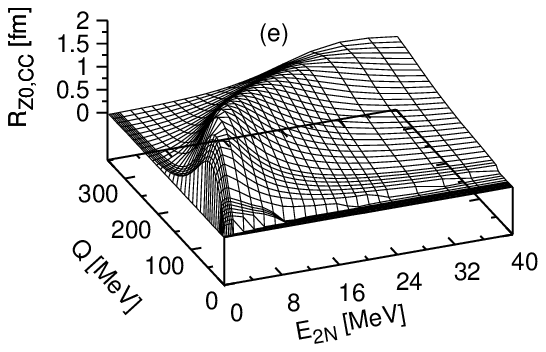}
\caption{The nuclear inclusive response functions 
$R_{00,CC}$ (a),
$R_{MM,CC}$ (b),
$R_{PP,CC}$ (c),
$R_{ZZ,CC}$ (d)
and
$R_{Z0,CC}$ (e)
for the $\bar{\nu}_e + {^2{\rm H}} \rightarrow e^+ + n + n$ reaction 
as a function of the internal 2N energy $E_{2N}$
and the magnitude of the three-momentum transfer $Q$.
The results are obtained with the AV18 NN potential
and the single nucleon CC operator,
which contains the relativistic corrections, employing 
the nonrelativistic kinematics.
}
\label{FIG.CC-ANTINUE-2H.3D}
\end{figure}

\begin{figure}
\includegraphics[width=0.45\textwidth]{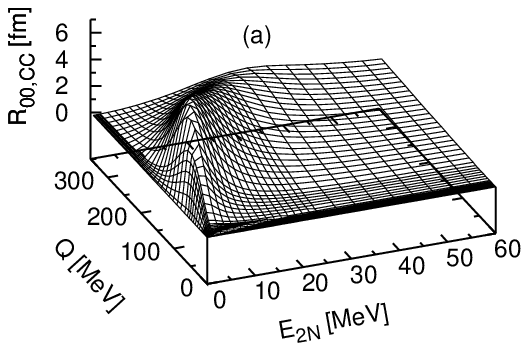}
\includegraphics[width=0.45\textwidth]{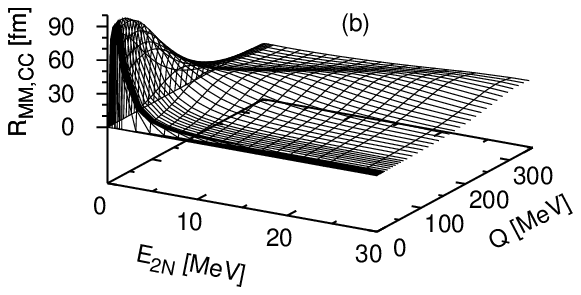}
\includegraphics[width=0.45\textwidth]{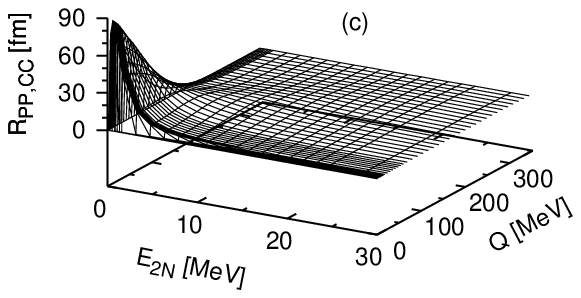}
\includegraphics[width=0.45\textwidth]{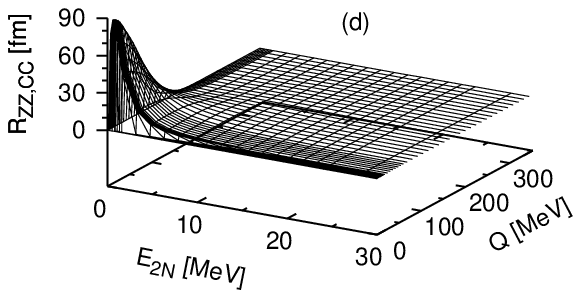}
\includegraphics[width=0.45\textwidth]{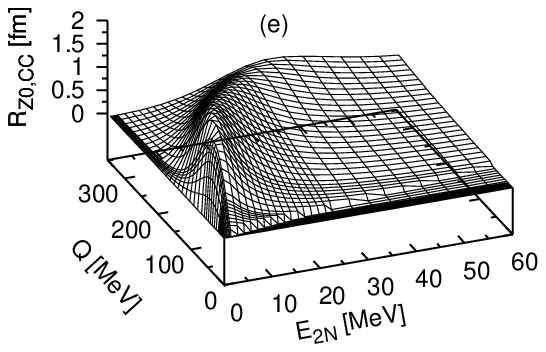}
\caption{The same as in Fig.~\ref{FIG.CC-ANTINUE-2H.3D} 
for the ${\nu}_e + {^2{\rm H}} \rightarrow e^- + p + p$ reaction.
}
\label{FIG.CC-NUE-2H.3D}
\end{figure}

\begin{figure}
\includegraphics[width=0.45\textwidth]{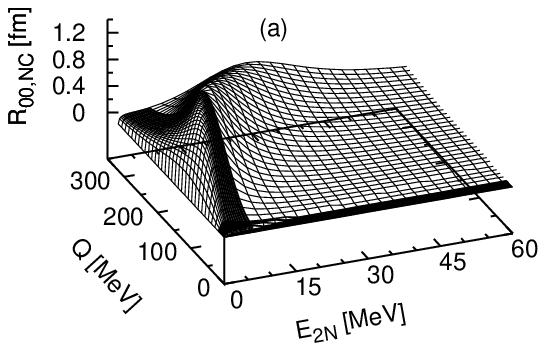}
\includegraphics[width=0.45\textwidth]{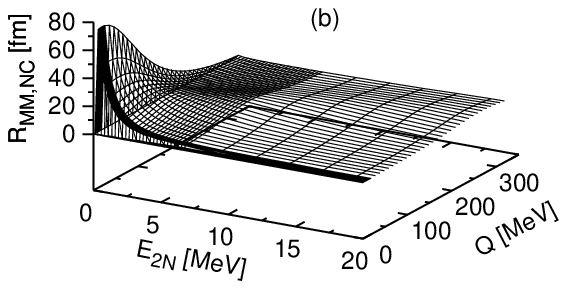}
\includegraphics[width=0.45\textwidth]{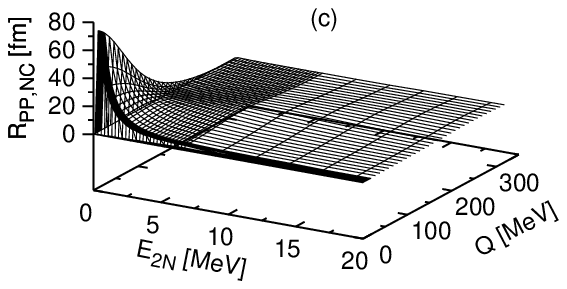}
\includegraphics[width=0.45\textwidth]{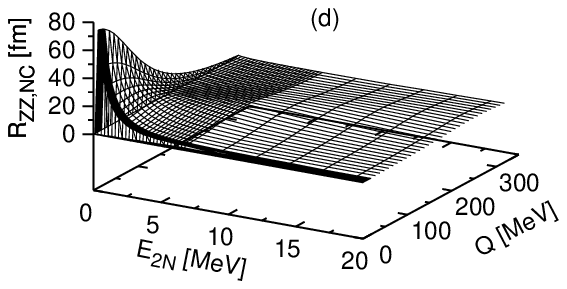}
\includegraphics[width=0.45\textwidth]{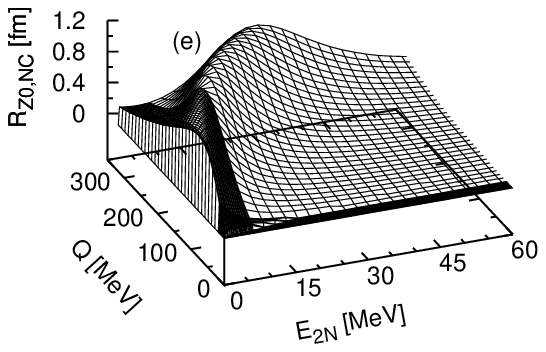}
\caption{The same as in Figs.~\ref{FIG.CC-ANTINUE-2H.3D} 
and \ref{FIG.CC-NUE-2H.3D} but for the weak NC driven 
$\bar{\nu}_e + {^2{\rm H}} \rightarrow \bar{\nu}_e + p + n$
and
${\nu}_e + {^2{\rm H}} \rightarrow {\nu}_e + p + n$ 
reactions.
}
\label{FIG.NC-2H.3D}
\end{figure}

In Figs.~\ref{FIG.CC-ANTINUE-2H.3D}--\ref{FIG.NC-2H.3D} 
we show the three sets of the response functions obtained 
for the $\bar{\nu}_e + {^2{\rm H}} \rightarrow e^+ + n + n$,
${\nu}_e + {^2{\rm H}} \rightarrow e^- + p + p$
and
${\nu}_e (\bar{\nu}_e) + {^2{\rm H}} \rightarrow {\nu}_e (\bar{\nu}_e) + p + n$ reactions.
We use different $E_{2N}$ and $Q$-ranges in the figures
to display the particular features
of the response functions. Note that the figures are {\em not} drawn with all calculated points,
so the actual grids for two dimensional interpolations are in fact much denser.
All the response functions have a maximum in the vicinity of the $(0,0)$ point but their shapes and heights
are very different. The response functions $R_{00}$ and $R_{Z0}$ stemming at least partly 
from the $N^0$ nuclear matrix elements
are, for all the three reactions, dominated by the response functions $R_{MM}$, $R_{PP}$ and $R_{ZZ}$,
which are by two orders of magnitude more pronounced.
Note that for each $E_{2N} > 0 $ there is an interval $[0,Q_{min}]$ which cannot be physically realized
for any initial neutrino energy and for which the values of the response functions are set to zero. 
The exact expression for $Q_{min}$ is quite complicated in the case when the final lepton is massive
so we give here only approximate expressions, assuming that the electron (or positron) mass can be neglected:
\begin{eqnarray}
Q_{min}= 2 \sqrt{{M} \left(-2 \sqrt{-{M} ({E_{2N}}-{M_d}+{M})}-{E_{2N}}+{M_d}\right)} \, .
\label{Qmin}
\end{eqnarray}
The corresponding maximal value of $Q$ for given $E_{2N}$ depends also on $E$:
\begin{eqnarray}
Q_{max}= 2 \sqrt{{M} \left(-2 \sqrt{{M} (2 E-{E_{2N}}+{M_d}-{M})}+2 E-{E_{2N}}+{M_d}\right)} 
\label{Qmax}
\end{eqnarray}
and the maximal value of $E_{2N}$ for given $E$ is obtained from the condition 
$Q_{min} = Q_{max}$ and reads
\begin{eqnarray}
( E_{2N} )^{max} = \frac{-E^2+4 E {M}+4 {M_d} {M}-8 {M}^2}{4 {M}} \, .
\label{E2NMAX}
\end{eqnarray}

From the response functions it is straightforward to compute the differential and total
cross sections. To this end one interpolates the response functions in two dimensions
over the $(E_{2N},Q)$ grid points to the particular $({\bar E}_{2N},{\bar Q})$ value 
resulting from the $(E, \theta, E^\prime \, )$ set. 
We used three different methods to interpolate the response functions. While the two first 
methods employed consecutive cubic 
splines (from Ref.~\cite{splines1} or from  Ref.~\cite{splines2}) interpolations,
first along the $Q$ direction and then along the $E_{2N}$ direction,
the third method was a straightforward bilinear interpolation. 
In this way we could control the quality of interpolations, since
we required that results for all considered observables,
obtained by the three methods, did not deviate from the average
by more than 1~\%. Additional points were added to the grid, when that criterion was not met.
Since the 2N calculations are relatively easy, we could consider grids which contained from 7200 to
17200 points. This procedure was especially important for the 3N case, where we did not calculate 
cross sections directly but fully relied on response functions' interpolations.
In the following we show results based on the interpolation scheme from Ref.~\cite{splines1}.

\begin{figure}
\includegraphics[width=0.31\textwidth]{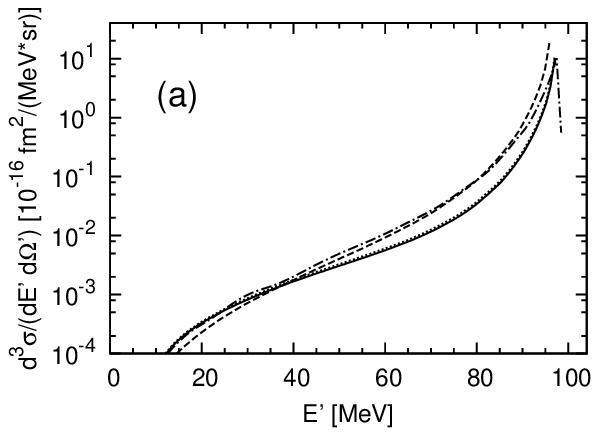}
\includegraphics[width=0.31\textwidth]{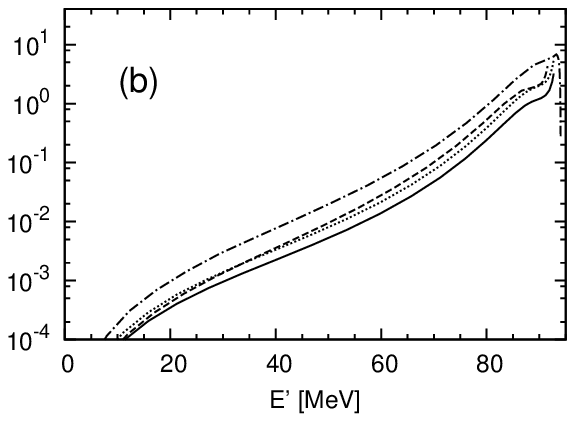}
\includegraphics[width=0.31\textwidth]{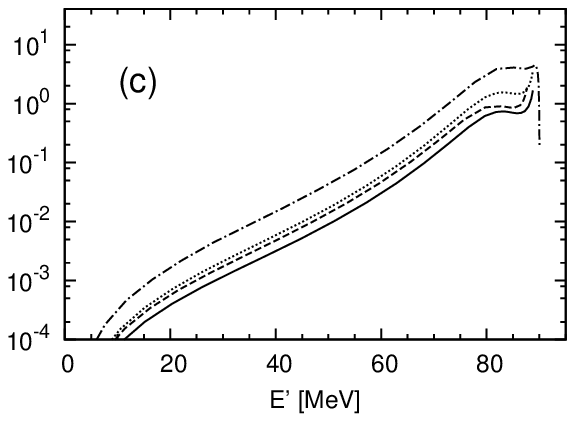}
\caption{The triple differential cross section 
$d^3{\sigma}/\left( dE^\prime \, d{\Omega}^\prime \, \right)$ for the
$\bar{\nu}_e + {^2{\rm H}} \rightarrow e^+ + n + n$ (dashed line),
${\nu}_e + {^2{\rm H}} \rightarrow e^- + p + p$ (dash-dotted line),
$\bar{\nu}_e + {^2{\rm H}} \rightarrow \bar{\nu}_e  + p + n$ (solid line)
and ${\nu}_e + {^2{\rm H}} \rightarrow {\nu}_e + p + n$ (dotted line) reactions
for the initial (anti)neutrino energy $E$= 100 MeV 
at three laboratory scattering angles: 
$\theta$= 27.5$^\circ$ (a),
$\theta$= 90$^\circ$ (b)
and
$\theta$= 152.5$^\circ$ (c)
as a function of the final lepton energy $E^\prime$. 
The results are obtained with the AV18 potential and with 
the single nucleon current, employing the nonrelativistic kinematics. 
\label{FIG.DBLE.2H}}
\end{figure}

The triple differential cross section,
$d^3{\sigma}/\left( dE^\prime \, d{\Omega}^\prime \, \right)$,
for a fixed lepton scattering angle 
is a function of the final lepton energy $E^\prime$. In Fig.~\ref{FIG.DBLE.2H} 
we show examples of such cross sections for the initial (anti)neutrino energy $E$= 100 MeV 
and for three different lepton scattering angles 
$\theta$= 27.5$^\circ$, $\theta$= 90$^\circ$ and $\theta$= 152.5$^\circ$.
These results can be compared with middle panels of Figs.~6 and~9 in Ref.~\cite{PRC86.035503}
and show that the cross sections rise very rapidly with the final (anti)lepton energy,
changing in the allowed energy range by several orders of magnitude.

\begin{figure}
\includegraphics[width=0.45\textwidth]{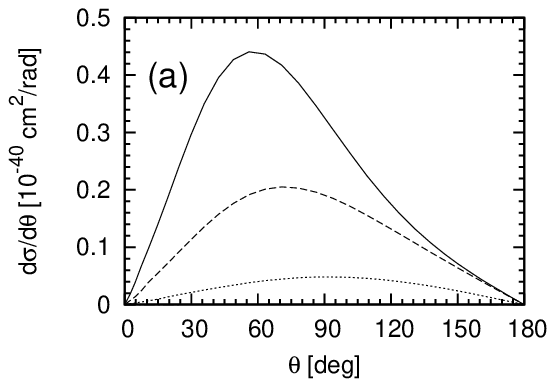}
\includegraphics[width=0.45\textwidth]{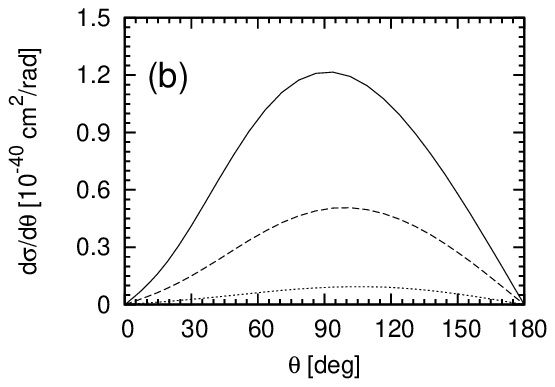}
\includegraphics[width=0.45\textwidth]{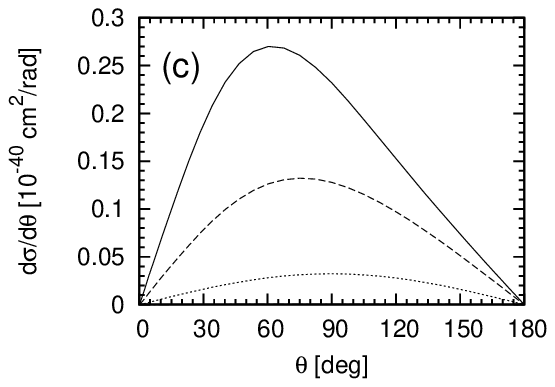}
\includegraphics[width=0.45\textwidth]{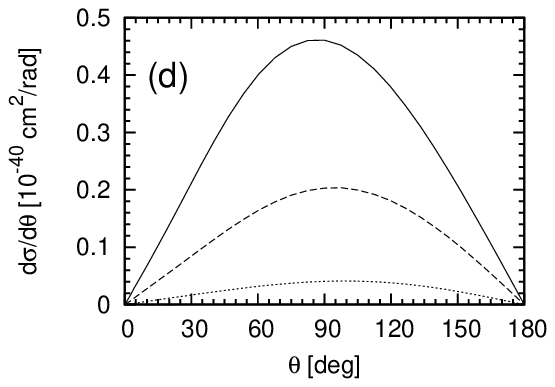}
\caption{The differential cross section $d{\sigma}/d{\theta}$ for the 
$\bar{\nu}_e + {^2{\rm H}} \rightarrow e^+ + n + n$ (a),
${\nu}_e + {^2{\rm H}} \rightarrow e^- + p + p$ (b),
$\bar{\nu}_e + {^2{\rm H}} \rightarrow \bar{\nu}_e  + p + n$ (c) 
and ${\nu}_e + {^2{\rm H}} \rightarrow {\nu}_e + p + n$ (d) reactions
as a function of the laboratory scattering angle $\theta$ 
for initial (anti)neutrino energy $E$= 50 MeV (dotted line), 
100 MeV (dashed line) and 150 MeV (solid line). 
The results are obtained with the AV18 potential and with 
the single nucleon current, employing the nonrelativistic kinematics. 
}
\label{FIG.ANG.2H}
\end{figure}

For all the four studied reactions we show also in Fig.~\ref{FIG.ANG.2H} the angular 
distributions of the cross sections, $d{\sigma}/d{\theta}$, which are given as
\begin{eqnarray} 
\frac{d{\sigma}}{d{\theta}} = 2 \pi \, \sin\theta \, \int\limits_{(E^\prime)_{min}}^{(E^\prime)_{max}}
dE^\prime \, \frac{ d^3{\sigma}}{ dE^\prime \, d{\Omega}^\prime \, } \, ,
\end{eqnarray} 
where $ {(E^\prime)_{min}} = M_e \, (0)$ for the CC (NC) induced reactions and the factor $2 \pi$ 
arises from the integration over the azimuthal angle $\phi$. 
Clearly, the angular distributions rise with the incident energy. For the smallest $E$= 50 MeV 
they are all almost symmetric with respect to $\theta$= 90$^\circ$. This symmetry is roughly preserved 
for the higher energies $E$= 100 and 150 MeV in the case of the two neutrino induced reactions but
for the two other reactions the angular distributions become asymmetric and their maxima 
are shifted towards forward angles. This behaviour is most evident 
for the $\bar{\nu}_e + {^2{\rm H}} \rightarrow e^+ + n + n$ process.

\begin{figure}
\includegraphics[width=0.95\textwidth]{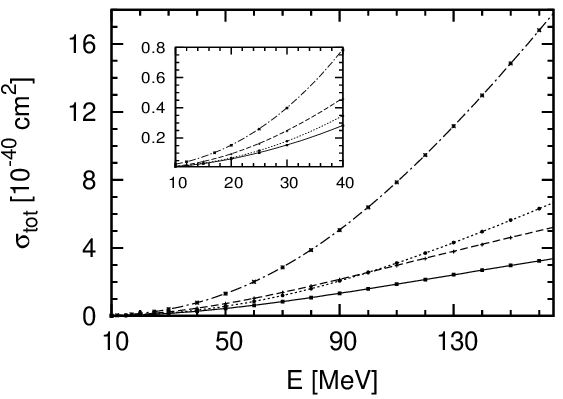}
\caption{The total cross section ${\sigma}_{tot}$ for the 
$\bar{\nu}_e + {^2{\rm H}} \rightarrow e^+ + n + n$ (dashed line),
${\nu}_e + {^2{\rm H}} \rightarrow e^- + p + p$ (dash-dotted line),
$\bar{\nu}_e + {^2{\rm H}} \rightarrow \bar{\nu}_e  + p + n$ (solid line) 
and ${\nu}_e + {^2{\rm H}} \rightarrow {\nu}_e + p + n$ (dotted line) 
reactions as a function of the initial (anti)neutrino energy $E$
calculated directly (symbols) or from the interpolated response 
functions (lines) as explained in the text.
The results are obtained with the AV18 potential and with 
the single nucleon current, employing the nonrelativistic kinematics.
The inset focuses on the results for $E \le$ 40 MeV. 
}
\label{FIG.TOT.2H}
\end{figure}

By the final integration over the scattering angle $\theta$ we arrive at the total 
cross section
\begin{eqnarray} 
\sigma_{tot} = \int\limits_0^\pi  d \theta \, \frac{d{\sigma}}{d{\theta} } \, .
\end{eqnarray} 
These important observables were presented 
in Refs.~\cite{PRC63.034617,NPA707.561,PRC86.035503,Baroni17}
and we remind the reader that our momentum space based results \cite{PRC98.015501}
agree very well with the predictions presented in \cite{PRC86.035503}. 
Despite the distinct treatment of kinematics the differences for 
none of the reactions for $E \le 150$ MeV exceed 2\% for
the single-nucleon current calculations and 6\% for
calculations including additionally two-nucleon currents.
In Fig.~\ref{FIG.TOT.2H} we display a comparison of directly obtained
results for the total cross sections with the predictions based on 
the response functions' interpolations. The agreement is very good for all the four 
reactions and for all considered (anti)neutrino energies, justifying our ``economical''
approach to calculations of the cross sections.

\section{Results for (anti)neutrino scattering on $^3$He and $^3$H}
\label{section5}

We follow the same path for the trinucleons as for the calculations 
with the deuteron target. That means that also in this case we 
calculate the response functions on a grid of $(E_{3N},Q)$ points. 
As already mentioned, one can evaluate the response functions 
by introducing explicit integrations over the available phase space,
in particular differentiating between the two- and three-body reaction
channels. It is also possible to evaluate the response functions
without any resort to explicit final-state kinematics \cite{incRAB,raport2005,PRC98.015501}. 
These two approaches were used and compared successfully in Ref.~\cite{PRC98.015501}
for a small number of $(E_{3N},Q)$ points.
Since we wanted to produce full grids of response functions, 
we decided to employ the second scheme. Each 3N grid comprised roughly 2000 points.
Even if some points on the rectilinear grids
lied in the nonphysical region, where
no calculations are necessary and where the response functions are just zero, the actual 
number of the computations was high. In order to efficiently deal with so many calculations,
we prepared a special computational framework to distribute the calculations 
among several desktop computers. Our calculations were performed with the AV18 NN potential,
neglecting the 3N force, and  
with the same single nucleon current as in the 2N case.
Since we do not include the proton-proton Coulomb force for the 3N scattering states, 
for the CC driven reactions we restricted ourselves to the antineutrino induced reactions,
which reduce the nuclear charge. We investigated also 
the NC reactions on $^3$He and $^3$H, although we are aware that 
our predictions for the weak NC response functions of $^3$He
might prove inaccurate for some parts of the phase space, where 
the proton-proton Coulomb force becomes important.

\begin{figure}
\includegraphics[width=0.45\textwidth,clip=true]{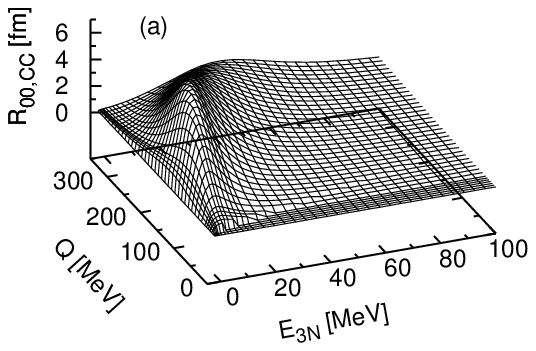}
\includegraphics[width=0.45\textwidth,clip=true]{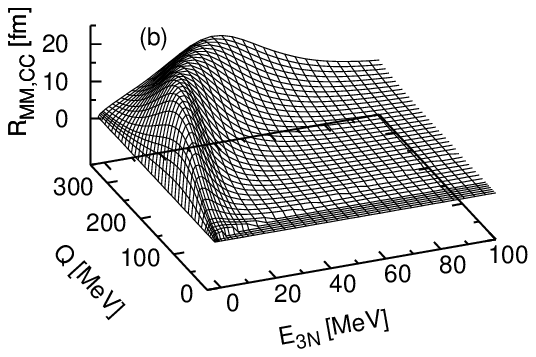}
\includegraphics[width=0.45\textwidth,clip=true]{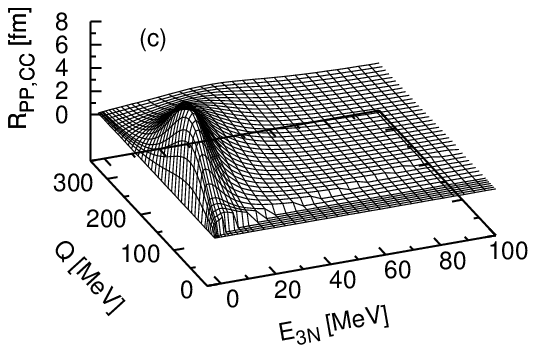}
\includegraphics[width=0.45\textwidth,clip=true]{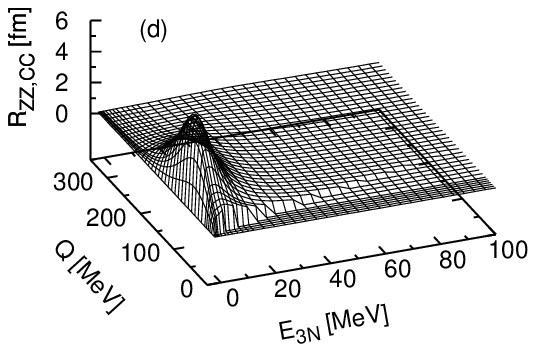}
\includegraphics[width=0.45\textwidth,clip=true]{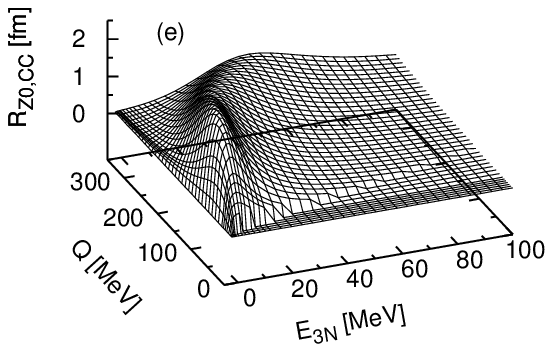}
\caption{The total inclusive CC response functions 
$R_{00,CC}$ (a),
$R_{MM,CC}$ (b),
$R_{PP,CC}$ (c),
$R_{ZZ,CC}$ (d)
and
$R_{Z0,CC}$ (e)
for the CC electron antineutrino disintegration of $^3$He
as a function of the internal 3N energy $E_{3N}$
and the magnitude of the three-momentum transfer $Q$.
The results are obtained with the AV18 NN potential
and the single nucleon CC operator,
which contains the relativistic corrections.
}
\label{FIG.CC-ANTINUE-3HE.3D}
\end{figure}

\begin{figure}
\includegraphics[width=0.45\textwidth,clip=true]{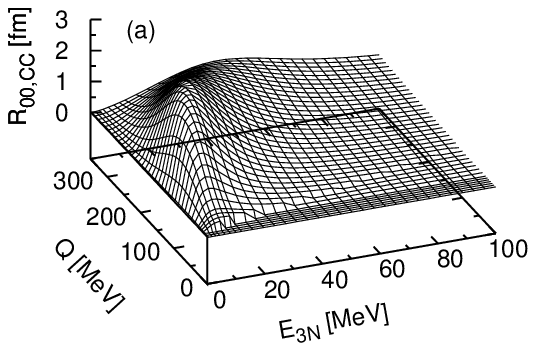}
\includegraphics[width=0.45\textwidth,clip=true]{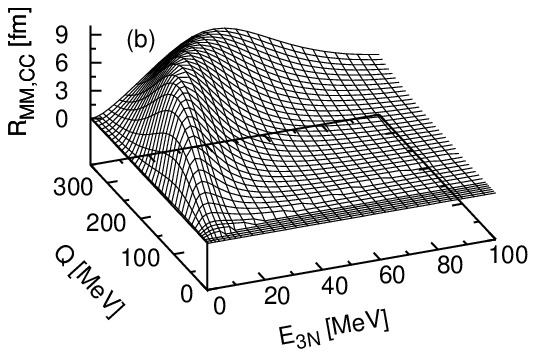}
\includegraphics[width=0.45\textwidth,clip=true]{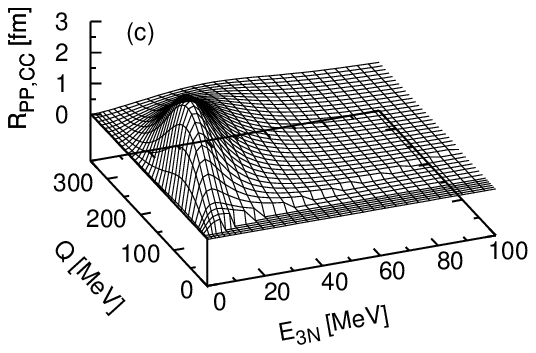}
\includegraphics[width=0.45\textwidth,clip=true]{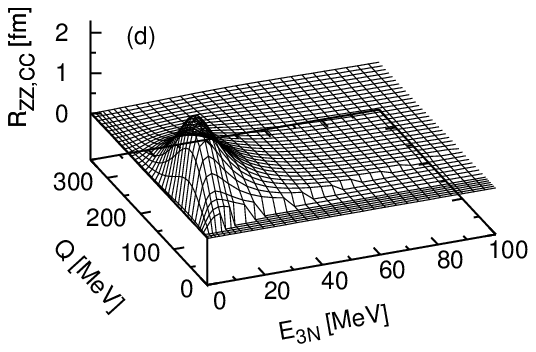}
\includegraphics[width=0.45\textwidth,clip=true]{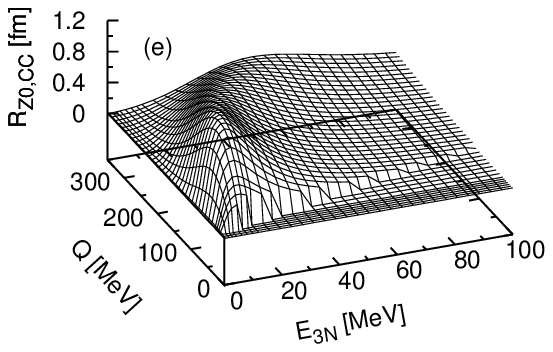}
\caption{The same as in Fig.~\ref{FIG.CC-ANTINUE-3HE.3D} but for the 
$\bar{\nu}_e + {^3{\rm H}} \rightarrow e^{+} + n + n + n$ reaction.
}
\label{FIG.CC-ANTINUE-3H.3D}
\end{figure}

\begin{figure}
\includegraphics[width=0.45\textwidth,clip=true]{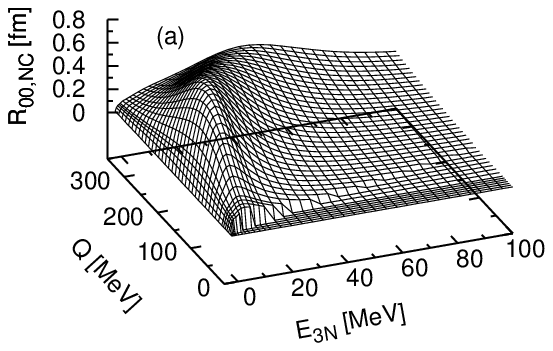}
\includegraphics[width=0.45\textwidth,clip=true]{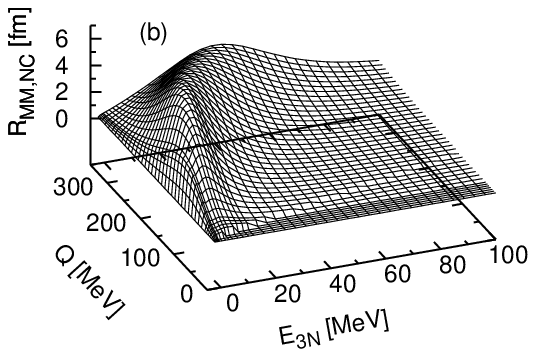}
\includegraphics[width=0.45\textwidth,clip=true]{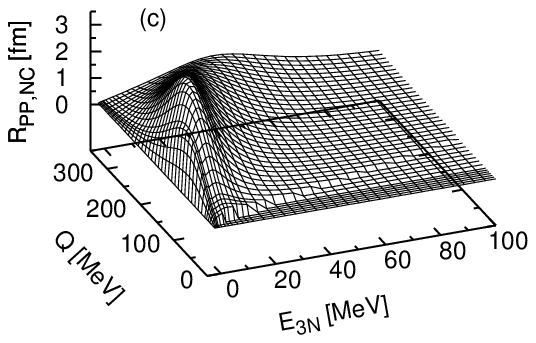}
\includegraphics[width=0.45\textwidth,clip=true]{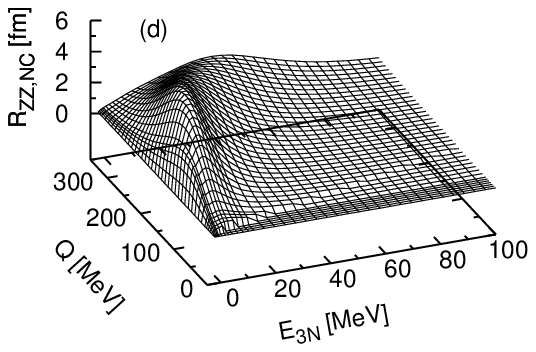}
\includegraphics[width=0.45\textwidth,clip=true]{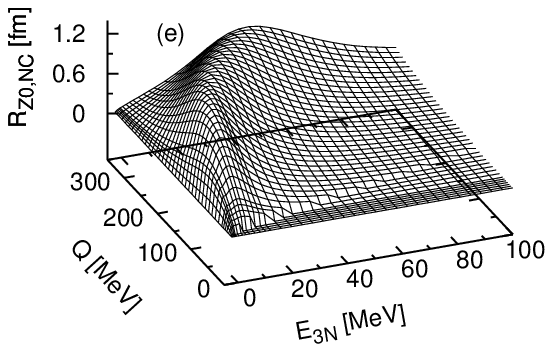}
\caption{The same as in Figs.~\ref{FIG.CC-ANTINUE-3HE.3D} 
and \ref{FIG.CC-ANTINUE-3H.3D} for
inclusive NC response functions
for (anti)neutrino disintegration of $^3$He.}
\label{FIG.NC-3HE.3D}
\end{figure}

\begin{figure}
\includegraphics[width=0.45\textwidth,clip=true]{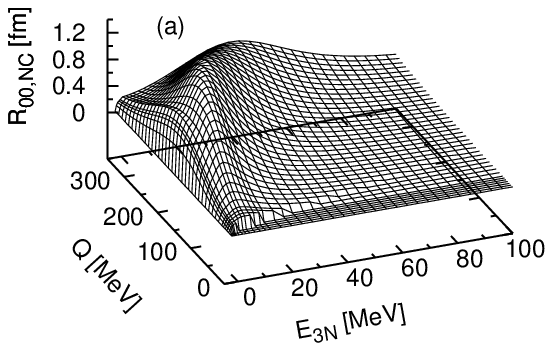}
\includegraphics[width=0.45\textwidth,clip=true]{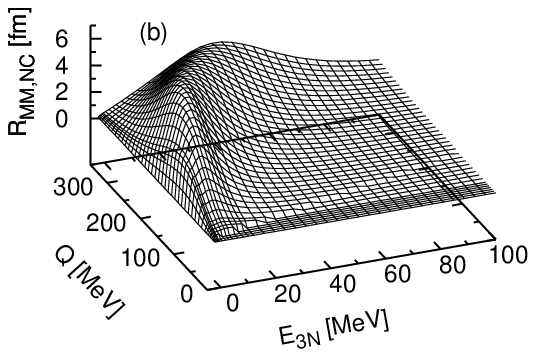}
\includegraphics[width=0.45\textwidth,clip=true]{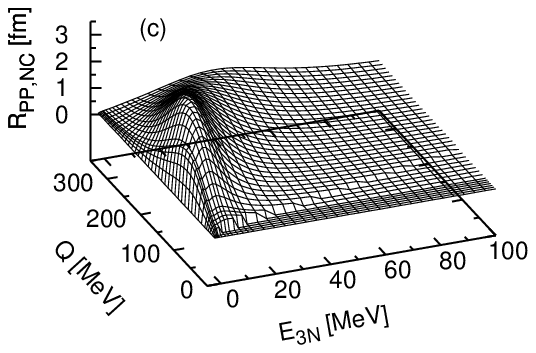}
\includegraphics[width=0.45\textwidth,clip=true]{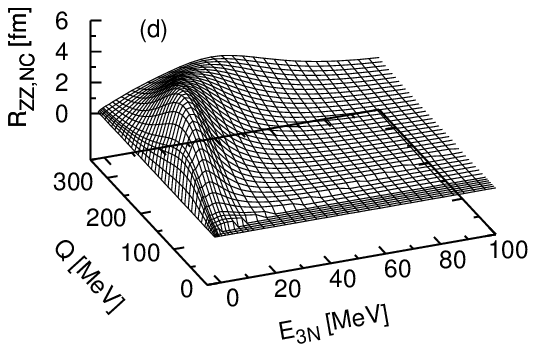}
\includegraphics[width=0.45\textwidth,clip=true]{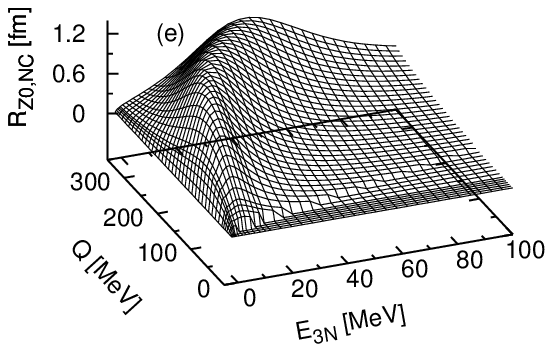}
\caption{The same as in Fig.~\ref{FIG.NC-3HE.3D} 
for inclusive NC response functions
for (anti)neutrino disintegration of $^3$H.}
\label{FIG.NC-3H.3D}
\end{figure}

We start presenting our results with the 3N weak response functions,
shown in Figs.~\ref{FIG.CC-ANTINUE-3HE.3D}--\ref{FIG.NC-3H.3D}.
It is clear that the 3N response functions are much broader and extend towards higher $E_{3N}$ 
and $Q$ values than the corresponding 2N observables, which are very localized. 
The differences between various response functions 
are not so strong as in the 2N case. The response functions
for the CC electron antineutrino disintegration of $^3$He and $^3$H (Figs.~\ref{FIG.CC-ANTINUE-3HE.3D}--
\ref{FIG.CC-ANTINUE-3H.3D})
have similar shapes and roughly scale according to the number of protons in a nucleus.
This seems to reflect the fact that the process described by the single nucleon current 
involves only protons.

In the case of the NC response functions 
the proton and neutron contributions 
to the single nucleon NC operator are comparable. This leads to similar
results for the $^3$He and $^3$H NC response functions displayed 
in Figs.~\ref{FIG.NC-3HE.3D}--\ref{FIG.NC-3H.3D}.

\begin{figure}
\includegraphics[width=0.31\textwidth]{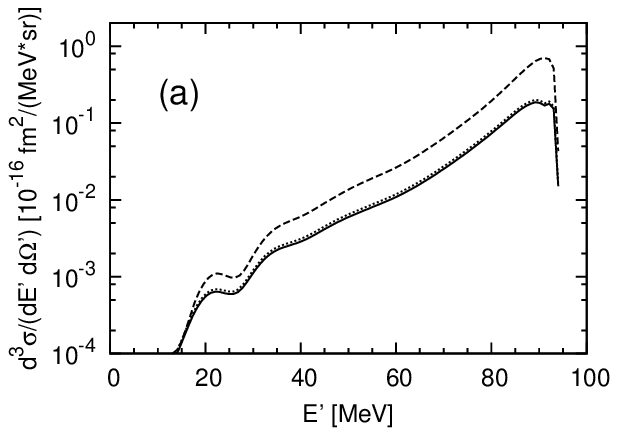}
\includegraphics[width=0.31\textwidth]{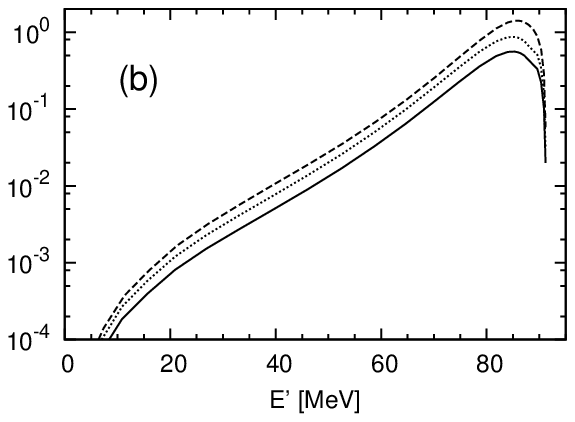}
\includegraphics[width=0.31\textwidth]{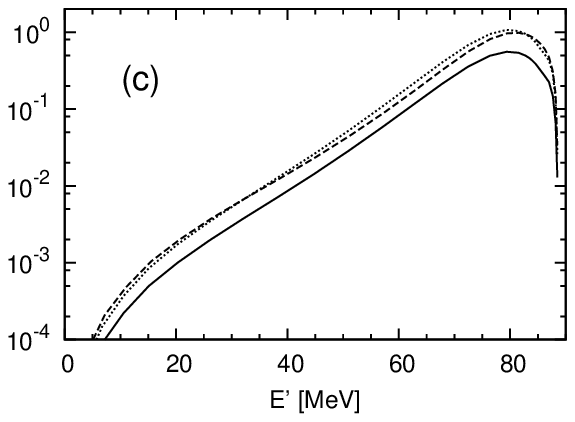}
\caption{The triple differential cross section
$d^3{\sigma}/\left( dE^\prime \, d{\Omega}^\prime \, \right)$ for the
CC electron antineutrino disintegration of ${^3{\rm He}}$ (dashed line), 
NC electron antineutrino disintegration of ${^3{\rm He}}$ (solid line)
and
NC electron neutrino disintegration of ${^3{\rm He}}$ (dotted line)
for the initial (anti)neutrino energy $E$= 100 MeV
at three laboratory scattering angles:
$\theta$= 27.5$^\circ$ (a),
$\theta$= 90$^\circ$ (b)
and
$\theta$= 152.5$^\circ$ (c)
as a function of the final lepton energy $E^\prime$.
The results are obtained with the AV18 potential and with
the single nucleon current, employing the nonrelativistic kinematics.
}
\label{FIG.DBLE.3HE}
\end{figure}

As in the 2N cases, the response functions are the key ingredients
of the cross sections, where only total or partial information 
about the final lepton is retained. 
The triple differential cross section
$d^3{\sigma}/\left( dE^\prime \, d{\Omega}^\prime \, \right)$ for the
CC electron antineutrino disintegration,
NC electron antineutrino disintegration,
and
NC electron neutrino disintegration of ${^3{\rm He}}$ just for one
initial (anti)neutrino energy $E$= 100 MeV
at three scattering angles $\theta$= 27.5$^\circ$, $\theta$= 90$^\circ$
and $\theta$= 152.5$^\circ$ are displayed in Fig.~\ref{FIG.DBLE.3HE}
as a function of the final lepton energy $E^\prime$.
All the cross sections soar with increasing $E^\prime$ and they are pulled down only in the vicinity 
of $(E^\prime)_{max}$.
At the selected forward angle the cross section for the CC driven process 
assumes the highest values and the results for the NC reactions with antineutrinos 
and neutrinos nearly overlap (at least observed on the logarithmic scale).
For the two other values of $\theta$ the antineutrino and neutrino NC cross sections are 
clearly separated and the cross section for the neutrino induced NC breakup of $^3$He 
is quite close to the prediction for the antineutrino induced CC process.

\begin{figure}
\includegraphics[width=0.31\textwidth]{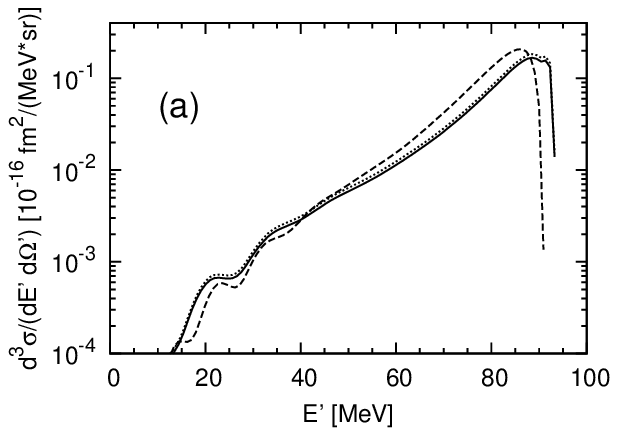}
\includegraphics[width=0.31\textwidth]{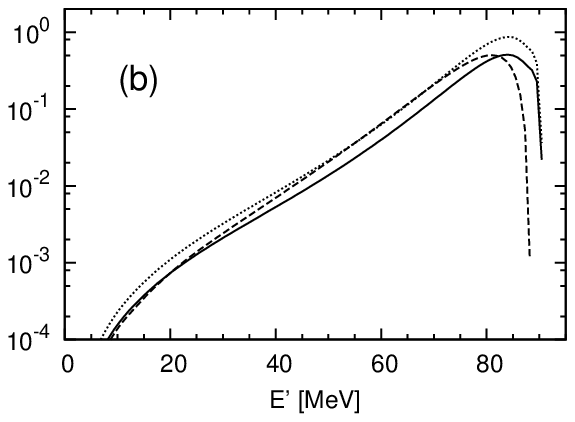}
\includegraphics[width=0.31\textwidth]{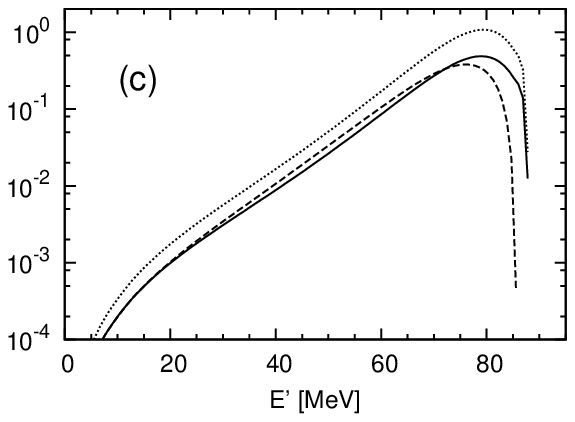}
\caption{The same as in Fig~\ref{FIG.DBLE.3HE} for the reactions
on ${^3{\rm H}}$.
}
\label{FIG.DBLE.3H}
\end{figure}

The corresponding predictions for the same reactions on $^3$H are shown in Fig.~\ref{FIG.DBLE.3H}.
At $\theta$= 27.5$^\circ$ the cross section for the CC reaction dominates for 
$ 40\, {\textrm MeV} <  E^\prime < 85 $ MeV but not over the whole 
$E^\prime$ interval. For $\theta$= 90$^\circ$ and $\theta$= 152.5$^\circ$ 
the cross section for the neutrino induced NC breakup of $^3$H assumes higher values
than the other two cross sections. 

We give only sample results but it is clear that similar calculations
can be used to plan experimental investigations 
of the NC and CC (anti)neutrino induced reactions.

\begin{figure}
\includegraphics[width=0.45\textwidth]{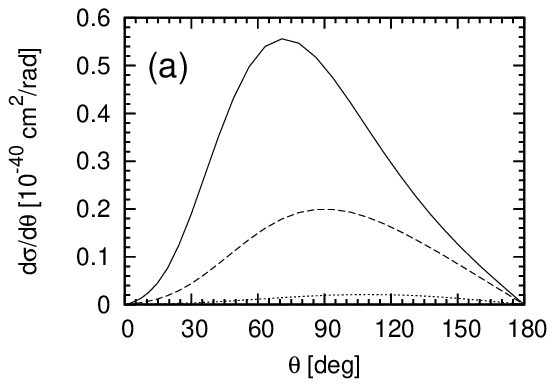}
\includegraphics[width=0.45\textwidth]{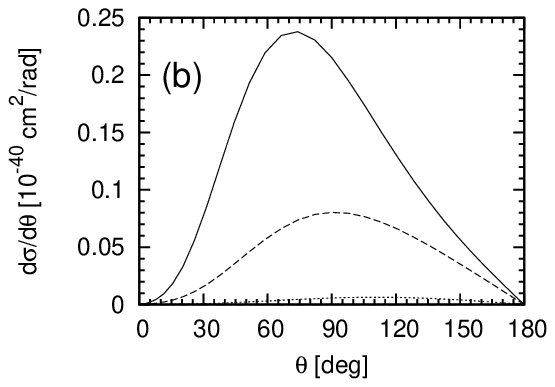}
\includegraphics[width=0.45\textwidth]{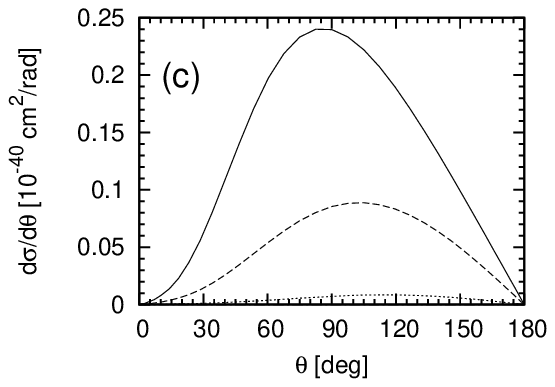}
\includegraphics[width=0.45\textwidth]{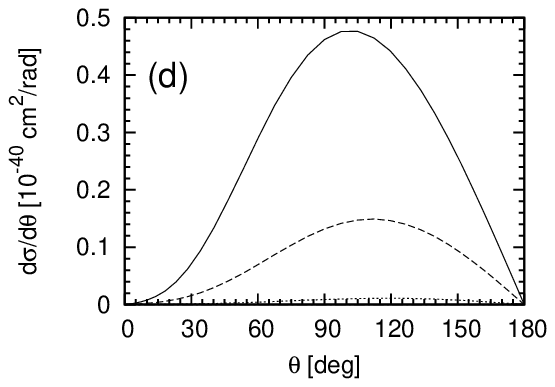}
\includegraphics[width=0.45\textwidth]{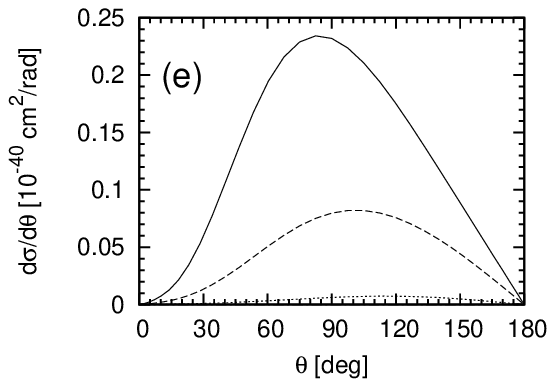}
\includegraphics[width=0.45\textwidth]{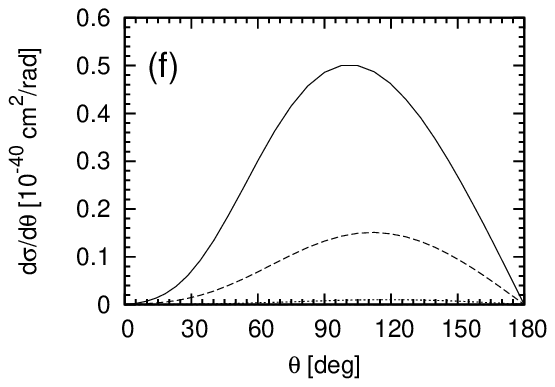}
\caption{The same as in Fig.~\ref{FIG.ANG.2H} for the inclusive 
CC electron antineutrino disintegration of ${^3{\rm He}}$ (a), 
CC electron antineutrino disintegration of ${^3{\rm H}}$ (b), 
NC electron antineutrino disintegration of ${^3{\rm He}}$ (c), 
NC electron neutrino disintegration of ${^3{\rm He}}$ (d), 
NC electron antineutrino disintegration of ${^3{\rm H}}$ (e)
and
NC electron neutrino disintegration of ${^3{\rm H}}$ (f).
}
\label{FIG.ANG.3N}
\end{figure}

As a last step before discussing the total cross section we show 
in Fig.~\ref{FIG.ANG.3N} the six angular distributions
of the cross sections, which can be now easily obtained from the response functions. 
We do it again for the same incoming (anti)neutrino energies as in Fig.~\ref{FIG.ANG.2H}
for the reactions on the deuteron. The curves are less symmetric compared to the 
predictions from Fig.~\ref{FIG.ANG.2H}, which is clearly visible for the two higher $E$ values.
There is a clear similarity between the results shown in parts (a) and (b) for the 
antineutrino induced CC processes on $^3$He and $^3$H, which can be traced back to the scaling properties 
of the corresponding response functions. For the highest energy $E$= 150 MeV the maxima for all the cross sections
with antineutrinos are shifted towards forward angles; only for the neutrino induced NC processes 
displayed in panels (d) and (f) the maxima are reached for $\theta > 90^\circ$.

\begin{figure}
\includegraphics[width=0.95\textwidth]{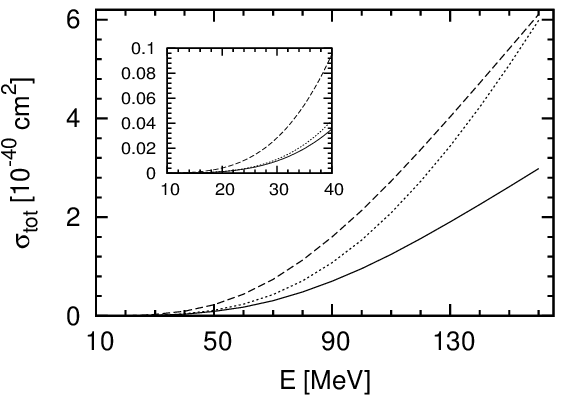}
\caption{The same as in Fig.~\ref{FIG.TOT.2H} for three inclusive 
(anti)neutrino reactions with $^3$He:
CC electron antineutrino disintegration of ${^3{\rm He}}$ (dashed line), 
NC electron antineutrino disintegration of ${^3{\rm He}}$ (solid line), 
NC electron neutrino disintegration of ${^3{\rm He}}$ (dotted line).
}
\label{FIG.TOT.3HE}
\end{figure}

\begin{figure}
\includegraphics[width=0.95\textwidth]{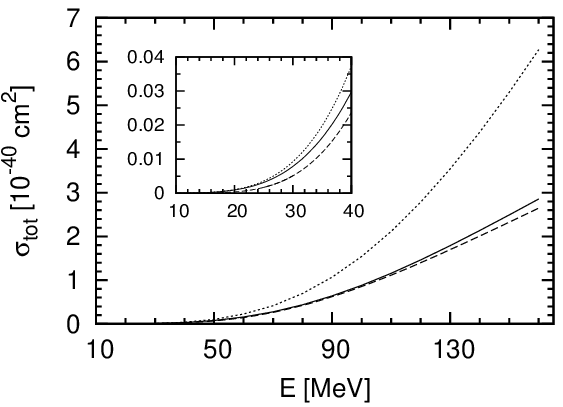}
\caption{The same as in Fig.~\ref{FIG.TOT.3HE} for three inclusive 
(anti)neutrino reactions with $^3$H.
}
\label{FIG.TOT.3H}
\end{figure}

Finally we arrive at the most important results - the total cross sections 
for the studied (anti)neutrino reactions with the trinucleons. 
They can be found in Fig.~\ref{FIG.TOT.3HE} for $^3$He and in Fig.~\ref{FIG.TOT.3H} for $^3$H.
In the $^3$He case the cross section for CC electron antineutrino disintegration 
takes the highest values in the whole investigated energy range. It is followed by the 
cross section for NC electron neutrino disintegration. The cross section 
for NC electron antineutrino disintegration is approximately two times smaller than 
the cross section for the corresponding CC driven process.

The picture is different for $^3$H, where the cross section for NC electron neutrino disintegration
is roughly two times larger than the predictions for the two antineutrino induced reactions, 
which are close to each other for all the initial (anti)neutrino energies.
 
\section{Summary}
\label{section6}

We extended our studies of (anti)neutrino scattering off the deuteron and trinucleons 
from Ref.~\cite{PRC98.015501}, where we presented our momentum space framework 
and obtained predictions for the total cross sections only for the reactions 
on the deuteron. For the reactions on the trinucleons we could 
perform in Ref.~\cite{PRC98.015501} only feasibility studies,
employing two different methods to calculate the essential response functions. 
In the present paper we provide information about the cross sections for 
CC electron antineutrino disintegration,
NC electron antineutrino disintegration
and
NC electron neutrino disintegration of ${^3{\rm He}}$ and ${^3{\rm H}}$.

The material presented in this paper is based on tens of thousands of 
2N and several thousands of 3N scattering calculations,
which were necessary to fill dense two dimensional grids, from which essentially 
in no time other observables: three fold differential cross sections, angular distributions
of the cross sections and, most importantly, the total cross sections can be obtained.
The results of our calculations in the form of the tabulated response functions 
are available to the interested reader.
This whole procedure was first carefully tested for the reactions on the deuteron, where the observables
had been calculated directly and where accurate predictions obtained in coordinate space
were available.

Our calculations leave room for improvement: 
they have been performed with the single nucleon current operator 
and without a 3N force, neglecting additionally the Coulomb force between
two final protons for one of the studied reactions. 
Nevertheless our predictions are obtained with the fully realistic AV18 nucleon-nucleon 
potential \cite{AV18} and are restricted to the (anti)neutrino energy region, where 
two-nucleon current and three-nucleon force effects are not expected to be very important 
and should not exceed 10~\%. Thus we provide important information about (anti)neutrino
interactions with very light nuclei.

A consistent framework
for the calculations of neutrino induced processes on
$^2$H, $^3$He, $^3$H and other light nuclei
is still a challenge, despite the recent progress in this field.
There are many models of the nuclear interactions and
weak current operators linked to these forces, but
full compatibility has not been achieved yet.
We hope that the work on the regularization of the 2N and 3N chiral potentials
as well as consistent electroweak current operators will be completed 
in the near future. This will allow us to repeat the calculations
of the response functions and related observables within a better 
dynamical framework. We believe, however, that the results 
presented in this paper constitute an important step 
towards a consistent framework
for the calculations of several neutrino induced processes on
$^2$H, $^3$He, $^3$H and other light nuclei.

\acknowledgments
This work is a part of the LENPIC project and was supported 
by the Polish National Science Centre under Grants
No. 2016/22/M/ST2/00173 and 2016/21/D/ST2/01120. The numerical calculations were
partially performed on the supercomputer cluster of the JSC, J\"ulich, Germany.

\end{document}